# Holographic Projection Technology: The World is Changing.

Ahmed Elmorshidy, Ph.D.

**Abstract--**This research papers examines the new technology of Holographic Projections. It highlights the importance and need of this technology and how it represents the new wave in the future of technology and communications, the different application of the technology, the fields of life it will dramatically affect including business, education, telecommunication and healthcare. The paper also discusses the future of holographic technology and how it will prevail in the coming years highlighting how it will also affect and reshape many other fields of life, technologies and businesses.

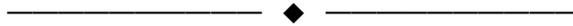

## 1  INTRODUCTION
### WHAT IS HOLOGRAPHIC PROJECTION?

Holographic projection is the new wave of technology that will change how we view things in the new era. It will have tremendous effects on all fields of life including business, education, science, art and healthcare. To understand how a holographic projector works we need to know what a hologram is. Holography is the method we use to record patterns of light. These patterns are reproduced as a three-dimensional image called a hologram. While Hungarian physicist Dennis Gabor invented the hologram in 1947. Today's new technology provides some outstanding advantages to not only everyday consumers but also large business corporations and governments. [10]

Three-dimensional holographic projection technology is loosely based on an illusionary technique called Peppers Ghost, and was first used in Victorian theatres across London in the 1860s. Pepper's Ghost was typically used to create ghostlike figures on stage. Hidden from the audience's view, an actor dressed in a ghostly costume would stand facing an angled plate of glass. The audience would be able to see the glass, but not the actor directly. [11]

_Dr. Ahmed Elmorshidy is with Gulf University for Science and Technology, Kuwait. Address: Gulf University for Science & Technology Block 5, Building 1,  Mubarak Al-Abdullah Area/West Mishref, Kuwait_
_Dr. Elmorshidy is also affiliated with Claremont Graduate University, California, USA._

3D holographic projection is a rapidly growing technology. With every business desperately trying to get their product to stand out from the competitors, 3D hologram advertising and promotion is fast becoming an eye catching success. Thanks to the latest in HD projection and CGI technology, 3D holographic projection has transformed itself from its basic Victorian origins into a futuristic audio visual display used by the likes of Endemol (Big Brother), Coco-Cola and BMW. With almost limitless holographic possibilities, from life like humans to blockbuster style special effects, as well as the continual advances in technology, 3D holographic projection has a bright future ahead. [15]

A holoprojector will use holographic technology to project large-scale, high-resolution images onto a variety of different surfaces, at different focal distances, from a relatively small-scale projection device. [13]  With many of the latest big budget cinema releases being available in 3D, and everyone talking about the 3D future of television, many eyes are starting to focus on 3D 1hologram projections without the glasses! [15]

## 2  IMPORTANCE AND NEED OF HOLOGRAPHIC PROJECTION

The interest in 3D viewing is not new. The public has embraced this experience since at least the days of stereoscopes, at the turn of the last century. New excitement, interest, and enthusiasm then came with the 3D movie craze in the middle of the last century, followed by the fascinations of holography, and most recently the advent of virtual reality. Recent developments in computers and computer graphics have made spatial 3D images more practical and accessible. The computational power now exists, for example, for desktop workstations to generate stereoscopic image pairs quickly enough for interactive display. At the high end of the computational power spectrum, the same technological advances that permit intricate object databases to be interactively manipulated and animated now permit large amounts of image data to be rendered for high quality 3D displays. [2]

Until currently, holographic data disks and holotechnology drives were just a matter of research. They were too costly and clumsy to use to be consumer markety feasible. However, recent improvements in the availability and cost reduction of lasers, digital cameras, and optical encoding substances are helping to turn the long-expected potential of holographic data storage into a commercial reality. The first holographic information disks were marketed consumer markety in the past year. Thus far, these holographic disks are still very costly and only Holographic Read Only Memory (HoloROM) is out. Nonetheless, rewritable holographic disks should come out in the next couple years. Further, manufacturing costs will decrease as product volume grows. This is the same configuration of improved product advancement and affordability that happened after CDs and DVDs were first launched. [9]

Modern three-dimensional ("3D") display technologies are increasingly popular and practical not only in computer graphics, but in other diverse environments and technologies as well. Growing examples include medical diagnostics, flight simulation, air traffic control, battlefield simulation, weather diagnostics, entertainment, advertising, education, animation, virtual reality, robotics, biomechanical studies, scientific visualization, and so forth. The increasing interest and popularity are due to many factors. In our daily lives, we are surrounded by synthetic computer graphic images both in print and on television. People can nowadays even generate similar images on personal computers at home. We also regularly see holograms on credit cards and lenticular displays on cereal boxes. [2]

There is also a growing appreciation that two-dimensional projections of 3D scenes, traditionally referred to as "3D computer graphics", can be insufficient for inspection, navigation, and comprehension of some types of multivariate data. Without the benefit of 3D rendering, even high quality images that have excellent perspective depictions still appear unrealistic and flat. For such application environments, the human depth cues of stereopsis, motion parallax, and (perhaps to a lesser extent) ocular accommodation are increasingly recognized as significant and important for facilitating image understanding and realism. [2]

In other aspects of 3D display technologies, such as the hardware needed for viewing, the broad field of virtual reality has driven the computer and optics industries to produce better stereoscopic helmet-mounted and boom-mounted displays, as well as the associated hardware and software to render scenes at rates and qualities needed to produce the illusion of reality. However, most voyages into virtual reality are currently solitary and encumbered ones: users often wear helmets, special glasses, or other devices that present the 3D world only to each of them individually. A common form of such stereoscopic displays uses shuttered or passively polarized eyewear, in which the observer wears eyewear that blocks one of two displayed images, exclusively one each for each eye. Examples include passively polarized glasses, and rapidly alternating shuttered glasses.

While these approaches have been generally successful, they have not met with widespread acceptance because observers generally do not like to wear equipment over their eyes. In addition, such approaches are impractical, and essentially unworkable, for projecting a 3D image to one or more casual passersby, to a group of collaborators, or to an entire audience such as when individuated projections are desired. Even when identical projections are presented, such situations have required different and relatively underdeveloped technologies, such as conventional autostereoscopic displays. Thus, a need still remains for highly effective, practical, efficient, uncomplicated, and inexpensive autostereoscopic 3D displays that allow the observer complete and unencumbered freedom of movement. Additionally, a need continues to exist for practical autostereoscopic 3D displays that provide a true parallax experience in both the vertical as well as the horizontal movement directions. [2]

A concurrent continuing need is for such practical autostereoscopic 3D displays that can also accommodate multiple viewers independently and simultaneously. A particular advantage would be afforded if the need could be fulfilled to provide such simultaneous viewing in which each viewer could be presented with a uniquely customized autostereoscopic 3D image that could be entirely different from that being viewed simultaneously by any of the other viewers present, all within the same viewing environment, and all with complete freedom of movement therein.Yet another urgent need is for an unobtrusive 3D viewing device that combines feedback for optimizing the viewing experience in combination with provisions for 3D user input, thus enabling viewing and manipulation of virtual 3D objects in 3D space without the need for special viewing goggles or headgear. In view of the ever-increasing commercial competitive pressures, increasing consumer expectations, and diminishing opportunities for meaningful product differentiation in the marketplace, it is increasingly critical that answers be found to these problems. Moreover, the ever-increasing need to save costs, improve efficiencies, improve performance, and meet such competitive pressures adds even greater urgency to the critical necessity that answers be found to these problems. [2]

## 3  APPLICATION OF HOLOGRAPHIC PROJECTION

With the use of the latest HD projectors, CGI animation, specialist HD film techniques and special effects created in post production, Pepper's Ghost technology has been upgraded to the 21st century. Instead of a real object or person's reflection appearing on a plate of glass, high definition video and CGI animation is beamed directly onto a specially designed, chemically treated transparent film via a high power HD projector. Although much more expensive, this modern approach results in a much clearer, believable hologram projection. [15]

In August 2009, Endemol, the producers of the famous reality TV show Big Brother, working together with activ8-3D holographic projections, beamed housemates' friends and families into the house to deliver messages of support and encouragement. The messages were pre-recoded using HD cameras and specifically angled lighting. A stage was rigged inside the Big Brother house task room, compiling of a HD projector, media player, lighting, and audio equipment. Each housemate entered the room in turn and took a seat in front of the stage. On cue, the housemate's family member or friend was beamed into the stage before delving their message. Although the hologram displays were difficult to judge on 2D television screens, the event was hailed as a great success, evoking brilliant reactions from the housemates which made for great TV. [15]

In January 2009 Coco-Cola gave a holographic sales conference presentation in Prague for over 800 people. Senior directors of the company were beamed into the stage as 3D holograms before giving a presentation about how the Coco-Cola brand has evolved over the years. The content of the presentation was also in the form of 3D holographic projections. The centre piece was a giant 3D hologram Coco-Cola branded spinning clock, representing the progression of time. A showcase of previous Coco-Cola bottles, logos, and labels amongst other objects were also projected as 3D holograms to create Prague's first 3D holographic projection display. [15]

3D Medical Animation Studio - 3d medical illustrations, has the capability of displaying 3d medical animations through holographic displays including the option of interactivity. Medical simulations company Tres 3d is pushing the boundaries of traditional MOA's (method of action) by creating holographic/3d animations to be viewed on holographic film without the need of special glasses. By using film with holographic properties and creating custom 3d computer medical animations Tres3d is able to create a holographic illusion. This process enables the audience to view the 3d medical animations with the illusion of depth. "Holographic Projection and 3d Imaging is nothing new to most of us. We all remember the first few 3d movies that required those silly paper glasses, picture a boardroom filled with top executives wearing them. I guess, that's why the technology never truly caught on."
Noted David Gonzalez, President of Tres3d Computer Medical Animation Studio (medical illustration studio) [17]

Conceptual medical illustrations and education illustration medical produced by Tres3d makes the top list of many conceptual medical illustration companies providing high quality 3d medical graphics. 3D illustration medical renders from the development of high quality models. Medical illustrator specialists from Tres 3d can create and render of any scientific visualization, illustration medical for high resolution print collateral. Human anatomy illustrations, medical graphics and conceptual medical illustration. 3d visualizations through mechanism of action illustrations. [14]

Conceptual medical illustration companies do not use global illumination technology when rendering 3d illustration medical. Tres 3d provides the most realistic renders available by using simulated natural light sources. Education illustrative medical are reviewed by Specialists on each 3d medical MOA. We have worked for companies like Johnson & Johnson, Glaxo Smith Kline and Unilever. Our clients and work speak volumes of our professionalism and high quality 3d graphics and medical artwork. Tres3d produces flash animation that loads quickly, DVD's to educate and train more effectively. Stills can be generated from each MOA method of action video for print materials. For 3d illustration medical campaigns of 3D conceptual medical illustrations models would be produced. [14]

A 3D holographic projection demonstration can be seen at The Movieum of London Museum, located in Westminster, England. The company behind it, activ8-3D holographic projections, are showcasing their large scale show and event holographic display, their medium and small size exhibition and retail hologram displays, as well as motion capture and interactive systems. For a free demonstration visit their website www.activ8-3d.co.uk or send an email to info@activ8-3d.com [15]

Real people can be filmed giving a speech, dance or presentation, for example, and then be projected as 3D holograms. [7]  Holographic special effects can be added in post-production to make a lifelike person beam into the room, Star-Trek style, or have their product appear and spin above their head at the click of their fingers.

In August 2009, Endemol - the producer of Big

Brother - working with Activ8-3D beamed housemates' friends and families into the house to deliver messages of support and encouragement. The messages were pre-recorded using HD cameras and specifically angled lighting. A stage was rigged inside the Big Brother house task room, comprising a HD projector, media player, lighting, and audio equipment. Each housemate entered the room in turn and took a seat in front of the stage. [7]

In January 2009, Coca-Cola gave a sales conference presentation in Prague for more than 800 people. Senior directors of the company were beamed into the stage as 3D holograms before giving a presentation about how the Coca-Cola brand has evolved over the years. The content of the presentation was also in the form of 3D holographic projections. The centrepiece was a giant 3D hologram Coca-Cola branded spinning clock, representing the progression of time. A showcase of previous Coca-Cola bottles, logos, and labels among other objects were also projected as 3D holograms. [7]

One can find many potential uses of holotechnology information systems in the general area of interactions and imaging. "HoloCams" will use holotechnology information storage and retrieval to store and virtually display 3D visual worlds. Holotech computer graphic interfaces and interfaces, such as gesture interpretation systems, will allow considerably more natural communication between a person and computer than can be done with present day two- dimensional image projections and keypad/mouse. Holotech imaging with temporal gated pulses will enable clear watching of things surrounded by light refracting matter such as body fluid or translucent atmospheres. Among the hindrances to widespread commercialization of holographic applied science is the price of integrated storage, processing, and display equipment, but prices of liquid crystal displays and digital camera chips are dropping. We are probably on the verge of widespread consumer marketization of holographic applied science. Sub-page toward lower cost holographic data storage covers additional useful information. [13]

With the use of the latest HD projectors, CGI animation, specialist HD film techniques and special effects created in post production, Pepper's Ghost technology has been upgraded to the 21st century. Instead of a real object or person's reflection appearing on a plate of glass, high definition video and CGI animation is beamed directly onto a specially designed, chemically treated transparent film via a high power HD projector. Although much more expensive, this modern approach results in a much clearer, believable hologram projection. [8] Due to the modern approach of projecting CGI animations and pre-recorded footage, almost anything is possible. The "blank canvas" approach is often adopted, creating a storyboard only limited by imagination. The storyboard can then be handed over to a CGI animation team who can make it come to life using the latest 3D software such as Maya or 3ds-Max. Real people can be filmed giving a speech, dance or presentation for example, and then be projected as 3D holograms. Holographic special effects can be added in post production to make a life-like person beam into the room, Star Trek style, or have their product appear and spin above their head at the click of their fingers. [8]

Bill Gates, Chairman of Microsoft Corp, made a virtual appearance at the "World Congress on Information Technology 2008", where he was reproduced on stage as a holographic simulation. The size of the projection was 4.6m and appeared in front of the audience of around 400 at the Kuala Lumpur Convention Centre. Apparently, the holographic image was very realistic and the crowd was impressed with the results. Gates stated during his speech that, "There are one billion people (in the world) who have a personal computer each but there are five billion others who don't. Microsoft also wants to reach these people."

After the flurry of 3D TVs we've seen over the past week or so, we've kind of got accustomed to the technology, but in China researchers have pushed those limits of 3D to another new level, we may still not be so ready for. The researchers in China have developed what they call the largest 3D holographic display. Measuring 1.8 × 1.3 m2 the screen is touted to be perfect for a great viewing experience with its continuous natural 3D image production. The 3D holographic display with the help of 64 digital cameras within is able to capture 3D images and with 64 projectors it is also able to recreate those images by projecting them on a holographic functional display screen at various angles. Because size of the magnification depends on the projector's ability and the camera array, there is no limit here on the holographic display, it can be used to create 3D images in any size, and still retain a high-quality image. The goal of the research team is to commercialize the display, but until then they expect to optimize the system and develop the requisite applications, software etc. [3]

In the area of marketing, holographic marketing or "holopromotion" is the application of holotechnology science to three-dimensional, high- resolution advertising. Multiple aspects of product marketing and purchasing might be put togetherd in bidirectional holograms that both draw people's attention and sell a product in real time. Holographically interactive booths and vending machines that can project consumer-responsive images several times their size will occupy much less space than traditional kiosks and vending machines. The site on telecommunications and holographic technology for more holotech information. [10]

Many holographic information storage systems have a beam splitter that splits a laser ray into two rays. One ray, known as the object beam, goes through a Spatial Light Modulator (SLM), like a LCD, that imprints data into the ray. The object ray intersects the second ray, known as the reference ray, inside optically sensitive recording media -- especially a photopolymer. The intersection of these rays, called an interference pattern, is a holographic image that is recorded in the media. When a reference beam is shown into the storage media at the same angle and wavelength used to store the holographic image, then the holographic image is recreated and the information configuration may be extracted and changed into electronic pulses. Data on holographic storage disks are expected to last as long as 50 years, while data on magnetic tape can erode in under a decade. For these reasons, uses for which archival durability is critical may increasingly switch from tape to holodisks. Linked page shape flexibility for holographic data storage also may be of interest to you. [10]

Holographic technology is already being used to projection television images onto a solitary pane of glass, giving the appearance of an image floating in mid-air. As the technology advances, true holographic television will likely project three dimensional images in mid-air without the need for a two dimensional projection surface. Investigators at research laboratories are working on applied science that could make holographic televisions (holovisions) that project moving, 3D pictures beyond the constraints of the device. True animated holography applied science lets people see travelling, three- dimensional pictures without the need for special glasses. One means of creating animated holographic pictures is to send laser light through a lithium niobate waveguide wrapped in piezoelectric material. A modulator converts video impulses into vibrations of the piezoelectric material that, consequently, alters the patterning of the light beam going through it. When this ray is projected into a translucent volume, it creates a 3D moving image. holographic technology and microelectronics deals with these concepts as well. [11]

Holograms are created by a pattern of overlapping peaks and troughs from the intersection of two rays of wave-synchronous laser light that come from dividing a laser source. Among the rays is known as the reference beam and the other is known as the signal ray. In holographic data storage, the signal ray contains information configurations that are written into optically sensitive media. The pattern of overlapping peaks and troughs formed by the intersection of these two beams contains the information and is imprinted into the optically sensitive media. That pattern, and therefore that information, can be retrieved later on by shining a reference laser ray into the media at the same angle that was used to burn the image. The site holographic data storage process provides additional information. [11]

The launch of Cisco's On-Stage TelePresence Experience, created by integrating Musion 3D Holographic Projection technology with Cisco TelePresence, was the result of an incredible effort on the part of 25+ Cisco employees across a half a dozen groups, plus another dozen individuals representing the vendor community. San Jose was jolted by an earthquake, informed De Beer. 'But we are fine, as you can see.' The 250 spectators nodded. Cisco's Telepresence has become `cuter', as Chambers put it, with holographic meetings replacing video conferencing in the near future. Cisco and Musion Systems will be marketing this as part of Cisco's Telepresence range of video conferencing products. Three-dimensional holographic conferencing will first be used at large expositions and conferences, and would later trickle down to enterprises. Over time, it might even be used at home. Your grandmother could virtually walk into a living room and talk to you - her image travelling over seas and countries over the Internet. A teacher could face 50 students and give a lecture complete with expressions and body language. The possibilities of this decidedly realistic application are numerous. [5]

Cellphones can use holographic technology in several ways. First, holographic technology allows the projection of an image that is several times the size of the projecting device without a projection screen. Second, this technology can allow virtual keyboards or other input devices that are larger than the device. One can find many potential applications of holographic data systems in the general area of interactions and imaging. "HoloCams" will employ holotechnology information recording and playback to store and project three-dimensional perceptual environments. Holotech computer image projections and interfaces, such as gesture recognition systems, will facilitate much more natural human-and-computer interaction than is feasible with present day two- dimensional displays and keyboard with mouse. Holographic imaging with temporal gated pulses will enable high-resolution views of things embedded in translucent material such as body fluid or translucent atmospheres. There is novel news at broad band sensory communication that may also be useful. [12]

In a recent development, with the use of 3D Holographic projection technology, apple has stated in a patent application they are they working on a cutting-edge model for a display that would use three-dimensional technologies. This seems relevant due to the fact that most recent developments in computer interface technology are starting to use this science in anything from entertainment to medical diagnostics. The effects of 3D technologies are immense in that human optics is adjusted specifically for three dimensional environments, so their use in practical not only in industry but for virtual reality applications as well. The pressure is on for manufacturers to implement this technology in products that would provide for the critical need for realistic holographic optics. What is surprising about the system is that it does not require the use of a 3D headset, instead uses autostereoscopic technology that adjusts as the viewer changes positions. In their patent they include the in-depth structure of the 3D rendering engine, and of course how the observer would interact with the image recognition abilities. [15]

Advertisers have been dreaming about three-dimensional video for half a century and marketing pundits have been predicting its coming for years. The applications for this technology are virtually limitless, but one thing is certain holographic technology captures the attention of those that see it. 360BrandVision™ introduces the only systems in the world that can display suspended holograms that can be seen 360 degrees and in any lighting environment for live events and entertainment. [18]

-- You have to see it to believe it. Advertisers have been dreaming about three-dimensional video for half a century and marketing pundits have been predicting its coming for years. The applications for this technology are virtually limitless, but one thing is certain holographic technology grabs the attention of those that see it. 360BrandVision™ currently offers the only systems in the world that can display suspended holograms that can be seen 360 degrees and in any lighting environment for live events and entertainment. [18]

On CNN election night "What people saw was the illusion of a life-size, video hologram that was created using post-production special effects. HD cameras and video software can be used to convincingly create a hologram that can be seen on television monitors, but not in real life", said Aaron Adjemian, 360BrandVision's VP of marketing. "360BrandVision's revolutionary Cheoptics360™ display systems actually let you create the illusion of 3D holographic video images suspended in midair." [18]

Another day, another patent, and if it was any company other than Samsung you could ask why they're concentrating on so much future tech instead of pushing phones for today out of the door. However Samsung had dozens of cellphones on show at the Mobile World Congress, and so I suppose feel quite justified playing around with 3D holographic displays; these would take advantage of the advances in rear-projection technology to not only produce a handset far thinner than is currently possible with traditional LCD panels but, by using a 3D hologram screen onto which the images are projected product, models that show depth in addition to height and width. [19]

The so-called "panel-type waveguide" is a refracting platter that spreads the projection out onto the display surface; it could be made in non-traditional shapes and sizes, giving manufacturers far more flexibility in cellphone design. [19]

Musion Systems have installed the UK's first 3D - Eyeliner™ Holographic projection system to appear in a retail mall as part of the new Toyota retail concept store launched at Bluewater Shopping Centre. Visitors can see a full sized car rotating in mid-air.

London, UK (PRWEB) January 30, 2007 -- Musion Systems, the projection technology behind the seminal video holograms of Sir Richard Branson, Madonna and virtual band Gorillaz, this week completed installation of the largest virtual 3D video hologram ever to appear in a single unit retail outlet - the projection of a full size Toyota Auris® family car spinning midair, complete with a life size human test driver. The groundbreaking projection, which dwarfs all previous in-store holograms, launches on January 18th for 4 weeks at the Bluewater shopping centre in Kent with Philips 3D plasma screens featuring at 7 other sites nationwide. The project was undertaken on behalf of Toyota GB in collaboration with creative agency Brandwidth Marketing and video production house Square Zero. [20]

Musion® Eyeliner™ is a new and unique high definition video projection system allowing spectacular freeform 3-dimensional moving images to appear within a live stage setting. Eyeliner brings dramatic, previously unseen 21st century video film effects to the live set mediums of stage, AV - artistic performances, conference or trade show presentations, retail displays and large scale out of home digital signage. [20]

The Musion installation is the main focal point of a unique series of in-store audio-visual experiences for the launch of Toyota's new car, the Auris. Creative agency Brandwidth Marketing together with leading retail design consultant John Whittle were commissioned by Toyota to design and build a unique retail theatre concept stores as a pivotal launch promotion for the Auris family hatchback car - a model seen by many as a serious challenger to the Ford Focus. The video presentations also include Philips WOWvx Intelligent 3D stereoscopic lenticular plasma screens provided by Musion Licensee partner Dimensional Studios. Both video mediums are able to present compelling virtual 3D images without the need for audiences to wear polarizing framed spectacles. [20]

Last year, Cisco and Munsion Systems performed the world's first real time 3D holographic telepresence video presentation. In other words, two men interacted live on a stage. But one was physically present in India, and the other was a life-size 3D holographic projection from California. Telepresence allows you to create live "in-person" experiences by combining telecommunications with advanced imagery. And when that imagery is 3D holographic technology, the experience is truly remarkable. Businesses use telepresence for meetings, and hospitals use it for training. But the areas of public speaking and presentation telepresence offer the greatest opportunities for churches. Obviously, this technology is very expensive, but perhaps one day it will quite affordable. For the latest on telepresence technology, visit Telepresence Options. According to ZDNet, by 2010 we will have holographic handsets that place 3D imaging capabilities in the hands of millions. [4]

# 4 FUTURE OF HOLOGRAPHIC POJECTION

3D holographic projection technology clearly has a big future ahead. As this audio visual display continues to get high profile credibility, we are likely to see more companies advertising their products or marketing their business in this way. Whether it be large scale, big budget product launches or smaller retail POS systems, they are likely to become a common feature in the advertising world. [15]

The holographic projectors that are under development will be able to be much smaller and portable than image projectors that rely on conventional, incoherent light beams. Ultimately, holographic projectors may become sufficiently small to be incorporated into future generation cell phones. Holographic techniques are being used for three-dimensional (3-D) rendering of medical pictures including MRI and CT pictures. Medical holotechnology imaging can enable doctors to test the insertion of medical instruments into an artificially constructed, three-dimensional version of the surgical field before the operation. An array of micro- mirrors, whose movements are controlled by computer, may be used to divide and focus an array of laser beams to make moving, three-dimensional holographic pictures of internal anatomic features. [9]

Holographic projectors will be able to render sharp projected images from relatively small projection devices (e.g. cell-phones) because they do not require high intensity, high-temperature light sources. Investigators at companies and universities are working toward applied science that could make television with holographic projections (holovisions) that can project moving, three-dimensional pictures outside of the screen. Genuine holographic motion picture science lets people see travelling, three-dimensional images absent special glasses. One means of creating animated holotechnology images is to send laser light by means of a lithium niobate waveguide covered by piezoelectric material. A modulator converts video impulses into vibrations of the piezoelectric material that, in turn, alters the configuration of the light beam going through it. When this ray is shone into a translucent volume, it creates a 3D travelling image. Two and three dimensional interference patterns presents some alternative perspectives. [10]

A holographic memory device that can store as much as five gigabytes could replace flash memory for many usages. It would be a boon to handheld machines like PDAs and smart phones. Next generation smart phones may use holotechnology applied science for data storage and display projection. For memory, holographic information recording and playback could significantly increase the memory capacity of phones. For display, holotechnology projection can show images, unconstrained by the tiny size of a handheld device. The idea of watching television on one's cellphone is in vouge now, but who wants to watch TV on a 2" screen? If it were possible to project a large picture from a cellphone onto a nearby wall, that would transform the use of cellphones for visual media. Also, storing data three-dimensionally with holographic storage has interesting notes on this holotechnology topic. [10]

In the areas of telecommunications and instruction, remote conferencing and distance education technologies featuring 2D screen pictures will evolve into three-dimensional, engaging holographic projection systems. Holographic applied science is, even now, being used for "HoloCells" (holographic cell phones that record and play three-dimensional, real time pictures of the communicating parties that may be viewed from different angles). The site three-dimensional medical imaging also provides information on these topics. [11]

A holographic memory device that can hold up to several gigabytes could compete with flash memory for several usages. It would be a boon to handheld machines especially smart phones and PDAs. Future versions of smart phones may use holotechnology applied science for both memory and display functions. For memory, holotechnology information recording and playback can greatly increase the memory capacity of phones. For display, holographic projection can show pictures that are not constrained by the small size of a mobile device. The prospect of watching TV on a mobile phone is in vouge now, but who wants to watch television on a 3" diagonal screen? If it were possible to project a large picture from one's cellphone onto the wall nearby, that would transform the use of cellphones for visual media. Linked page media for holographic data storage also has different information on this topic. [11]

Holographic applied science can also create new methods for three- dimensional visual contact from computing systems to human beings. This starts with screen displays with improved 3D projection qualities and then improves to mid-air, three- dimensional computer projections that do not require a screen. Similar holotech coverage at holographic technology and navigation may be of interest. [12]

Design is central to applied science, new product development and model building, building design and construction, pharmaceuticals, biologicals, and nanopharmacology, biochemistry and modeling at the molecular scale, biomedical technology and prostheses, the apparel industry, the fine arts, and other areas as well. Holotech applied science can help design for: manipulation of 3D models of molecules or biological structures; assembling electronics; and other design-related tasks. Linked page 3D imaging using micro-mirror arrays also deals with these technologies. [12]

The quantity of realized and potential usages of holographic science in the area of interpersonal interactions is also increasing quickly. A holographic camera (holocam) records and conveys radial three-dimensional real-time pictures from a central point using holographic applied science. A holoviewer projects these images for viewing in another location. Holocams and holographic viewers will probably be integrated into internet access, television, and cell phones in the next ten years. New telecommunication networks built on holotechnology science may be developed with uses in both personal and business interactions. Holographic science may also enhance the transferral speed and channel capacity for interactions systems based on fiber optics. To continue on related topics, see also holographic communication between humans and computers . [12]

At the present time, DVDs and CDs are still the main formats for mobile information storage media for music, video, and information. These traditional data storage media store information as distinct bits on the surface of the recording medium and the medium should be spun around to recover the information. The price of saving information is dropping but the need, however, for long-term information storage has been increasing even more promptly. Holotech information storage opens possibilities for saving information at much higher densities than CDs and DVDs by storing information three- dimensionally throughout the thickness of the recordable media. Visit also holographic data storage process . [13]

It sounds a lot like a wacky dream, but don't be surprised if within our lifetime you find yourself discarding your plasma and LCD sets in exchange for a holographic 3-D television that can put Cristiano Ronaldo in your living room or bring you face-to-face with life-sized versions of your gaming heroes. The reason for renewed optimism in three-dimensional technology is a breakthrough in rewritable and erasable holographic systems made earlier this year by researchers at the University of Arizona. Dr Nasser Peyghambarian, chair of photonics and lasers at the university's Optical Sciences department, told CNN that scientists have broken a barrier by making the first updatable three-dimensional displays with memory. [16]

According to Peyghambarian, they could be constructed as a screen on the wall (like flat panel displays) that shows 3-D images, with all the image writing lasers behind the wall; or it could be like a horizontal panel on a table with holographic writing apparatus underneath.

So, if this project is realized, you really could have a football match on your coffee table, or horror-movie villains jumping out of your wall. Peyghambarian is also optimistic that the technology could reach the market within five to ten years. He said progress towards a final product should be made much more quickly now that a rewriting method had been found.

However, it is fair to say not everyone is as positive about this prospect as Peyghambarian.

Justin Lawrence, a lecturer in Electronic Engineering at Bangor University in Wales, told CNN that small steps are being made on technology like 3-D holograms, but, he can't see it being ready for the market in the next ten years. "It's one thing to demonstrate something in a lab but it's another thing to be able to produce it cheaply and efficiently enough to distribute it to the mass market," Lawrence said. Yet, there are reasons to be optimistic that more resources will be channeled into developing this technology more quickly.

The Japanese Government is pushing huge financial and technical weight into the development of three-dimensional, virtual-reality television, and the country's Communications Ministry is aiming at having such technology available by 2020. Peyghambarian said there are no major sponsors of the technology at present, but as the breakthroughs continued, he hopes that will change. [16]

**Dr. Ahmed Elmorshidy** received his Ph.D. in Management of Information Systems (MIS) in 2004 from Claremont Graduate University, Claremont, California, U.S.A. previously he has earned an MBA in 1995 and an M.A. in Computer Resources and Information Management in 1994 from Webster University, St. Louis, Missouri, U.S.A. Dr. Elmorshidy's B.S. degree was in business administration from Alexandria University, Egypt. Dr. Elmorshidy taught at several academic institutions including Alexandria University, Webster University, Claremont Graduate University, National University, and currently at Gulf University for Science and Technology in Kuwait. Dr. Elmorshidy's research interests are focused on online information systems and the effect of new and disruptive technologies on the field of MIS (Management of Information Systems). Dr. Elmorshidy is member of IEEE organization and in is the Association of Information Systems (AIS).